\documentclass[a4paper, 12pt]{article}
\usepackage[bottom=2cm, top=1.5cm, left=2cm, right=2cm]{geometry}
\usepackage{amsmath,amssymb,amsfonts,amsthm,hyperref,bbm,latexsym,slashed}
\usepackage{graphicx,multirow, multicol}
\usepackage{cases}
\usepackage{cancel}
\usepackage{url}
\usepackage{indentfirst}
\usepackage[compat=1.1.0]{tikz-feynman}
\usepackage{gensymb}
\usepackage{subfigure}
\usepackage{orcidlink}
\usepackage{authblk}
\usepackage{tikz}
\usepackage[width=\textwidth]{caption}
\usepackage[english]{babel}
\usepackage{csquotes}
\usepackage[
backend=biber,
defernumbers=false,
citestyle=numeric-comp,
bibstyle=chem-angew,
sorting=none,
sortcites=true,
doi=true,
eprint=true, 
url=true, %DOI and URL seem redundant, so I turned URL off.
giveninits=true
]{biblatex} 
\allowdisplaybreaks

\newcommand\sfrac[2]{{\textstyle \frac{#1}{#2}}}

\begin{document}

\title{Decays of heavy scalars in the Grimus--Neufeld model}

% \author{
%   Name Surname~\orcidlink{0000-0000-0000-0000}\thanks{E-mail:
%     \href{mailto:name.surname@ff.vu.lt}{name.surname@ff.vu.lt}}
%   \\
%   \small Vilnius University, Faculty of Physics, Institute of Theoretical Physics and Astronomy, \\
%   \small Saulėtekio av. 9, Vilnius, Lithuania, LT-10222
% }

%% SD: added a package for authors and ordered our names alphabetically, so that Aurimas is the first author :)
\author{Aurimas Vitkus~\orcidlink{0009-0006-8604-9180} 
        \thanks{E-mail:\href{mailto:aurimas.vitkus@ff.stud.vu.lt}{aurimas.vitkus@ff.stud.vu.lt}}
        }
\author{Simonas Draukšas~\orcidlink{0000-0003-4796-2760}
        \thanks{E-mail:\href{mailto:simonas.drauksas@ff.vu.lt}{simonas.drauksas@ff.vu.lt}}
        }
\author{Thomas Gajdosik~\orcidlink{0000-0002-4355-8878}
        \thanks{E-mail:\href{mailto:tgajdosik@yahoo.com}{tgajdosik@yahoo.com}}
        }
\affil{\small Vilnius University, Faculty of Physics, Institute of Theoretical Physics and Astronomy, \\ Saulėtekio av. 9, Vilnius, Lithuania, LT-10222}

\date{}

\maketitle

\begin{abstract}
We consider an extension of the Standard Model by an additional Higgs doublet and a Majorana neutrino, which we call the Grimus--Neufeld Model (GNM). For certain parameter choices the GNM can be compared to the Inert Doublet Model (IDM), which has a scalar dark matter candidate. This motivates that the scalars of the GNM could possibly contribute to dark matter. To check this, we present the tree-level two-body decays of the heavy scalars of the GNM and compute the lifetime of the pseudoscalar in the IDM limit.
\end{abstract}

%\pagebreak
\tableofcontents

\section{Introduction}
The Standard Model (SM) of particle physics has passed numerous experimental tests~\cite{cms2025,atlascollaboration2024} and is the most successful model so far. However, it fails to explain neutrino oscillations~\cite{fukuda1998} and dark matter (DM)~\cite{cirelli2026}. Dark matter candidates have to be electromagnetically neutral and their lifetimes have to be longer than the age of the universe. Many models try to explain dark matter by introducing additional particles, that fit these conditions. Inert Doublet Models (IDM) extend the scalar sector, e.g.~\cite{deshpande1978,ma2006,krawczyk2016}. Other models extend the fermion sector, for example, by adding a sterile neutrino, e.g.~\cite{dodelson1994,ghiglieri2015,drewes2016,boyarsky2019,rojas2019}. 

The Grimus--Neufeld model (GNM)~\cite{grimus1989,dudenas2022,drauksas2024} extends the SM by adding a second Higgs doublet and one sterile neutrino that mixes with the active neutrinos. At tree-level the model has two massless neutrinos and two massive ones coming from the well known seesaw mechanism~\cite{mohapatra1980,schechter1982,giganti2018}. At $1$-loop level one of the massless neutrinos gets a mass via radiative corrections involving the second Higgs doublet.

The additional neutral particles of the model, i.e. the sterile neutrino and the neutral Higgs bosons, are massive and interact weakly, making them weakly interacting massive particles (WIMPs). To be proper dark matter candidates their lifetimes have to be longer than the age of the universe.

In this paper we estimate lower bounds on the lifetimes of the additional neutral Higgs bosons $H_0$ and $A$ by calculating all possible two-body tree-level decays. As a lower limit for the mass of the additional Higgs bosons we take $60~\mathrm{GeV}$, corresponding to the usual limit found in inert doublet models, e.g.~\cite{krawczyk2016}.

This paper is organized as follows: 
in Section~\ref{sec:gnmlagrang} we introduce the Grimus--Neufeld model, in Section~\ref{sec:Higgsdecays} we present the tree-level Higgs decays, in Section~\ref{sec:GNDM} we discuss our results, and we give our conclusions in Section~\ref{sec:Conclusions}.

\section{The Grimus--Neufeld Lagrangian} \label{sec:gnmlagrang}

In this section we write down the Lagrangian of the Grimus--Neufeld model (GNM) that will be used for our calculations. As an extension of the Standard Model (SM), the GNM uses all fields in the standard representations of the gauge groups of the SM. We are using latin letters from the beginning of the alphabet for the gauge group indices of the adjoint representation and for sums over Higgs doublets, latin letters from the middle of the alphabet for the gauge group indices of the fundamental representation and for the generation index, and greek letters for space-time indices, similar to~\cite{romao2012}. Since we restrict ourselves to one loop accuracy, we can ignore quarks and gluons in this paper.

For consistency we repeat some of the definitions to make our conventions better visible. 
The gauge covariant derivative follows the convention of FeynRules~\cite{alloul2014}, i.e. picking the phases $\eta$ and $\eta^{\prime}$ from \cite{romao2012} as $-1$, 
\begin{equation}
  D_{\mu}
= 
  \partial_{\mu} 
- ig W_{\mu}^{a} T^{a}
- ig^{\prime} B_{\mu} Y
\enspace ,
\label{eq:covderiv}
\end{equation}
with generators $T^{a}$ for the gauge group $SU(2)_{L}$ and generator $Y$ for the gauge group $U(1)_{Y}$ gives the non-Abelian field strength tensor
\begin{equation}
  W^{a}_{\mu\nu}
=
  \partial_{\mu} W^{a}_{\nu}
- \partial_{\nu} W^{a}_{\mu}
+ g \epsilon^{abc} W^{b}_{\mu} W^{c}_{\nu}
\label{eq:wmunu}
\end{equation}
and the Abelian field strength tensor
\begin{equation}
  B_{\mu\nu}
=
  \partial_{\mu} B_{\nu}
- \partial_{\nu} B_{\mu}
\enspace .
\label{eq:bmunu}
\end{equation}
These are used for the Lagrangian of the gauge sector
\begin{equation}
  \mathcal{L}_{\text{gauge}}
=
- \sfrac{1}{4} W^{a}_{\mu\nu} W^{a\,\mu\nu}
- \sfrac{1}{4} B_{\mu\nu} B^{\mu\nu}
\enspace , 
\label{eq:gauge}
\end{equation}
where the sum over the gauge index $a$ is implied.

Writing the fermionic doublets 
\begin{equation}
  L_{k}
=
  \begin{pmatrix} \nu_{k} \\ \ell_{Lk} \end{pmatrix}
\label{eq:leptdoubl}
\end{equation}
with capital letters we get the Lagrangian of the fermion-gauge sector 
\begin{equation}
  \mathcal{L}_{\mathrm{F-G}}
=
  \sum_{k = \text{generation}} 
  \bar{L}_{k} i \slashed{D} L_{k}
+ \bar{\ell}_{R k} i \slashed{D} \ell_{R k}
\enspace ,
\label{eq:gaugeferm}
\end{equation}
using the Feynman-slash notation 
\begin{equation}
  \slashed{D} := \gamma^{\mu} D_{\mu}
\enspace . 
\label{eq:Feynman-slash}
\end{equation}

For the Higgs sector we keep close to the conventions of~\cite{branco2012,gunion2018}, but go immediately to the Higgs basis~\cite{haber2011}
\begin{equation}
  H_{1} 
=
  \begin{pmatrix}
    \phi^{+}
  \\
    \frac{1}{\sqrt{2}}(v+h_{1}+i\phi^{0})
  \end{pmatrix}
\enspace , \quad
  H_{2} 
=
  \begin{pmatrix}
    h_{2}^{+}
  \\
    \frac{1}{\sqrt{2}}( h_{2} + i h_{3} ) 
  \end{pmatrix}
\enspace .
\label{eq:1sthiggs}
\end{equation}
Then the Higgs Lagrangian consists of the kinetic terms and the Higgs potential 
\begin{equation}
  \mathcal{L}_{\mathrm{H}}
=
  \sum_{a=1}^{2}
  ( D_{\mu} H_{a} )^{\dagger} ( D^{\mu} H_{a} )
- V(H_{1},H_{2})
\enspace , 
\label{eq:higgs}
\end{equation}
where we follow~\cite{drauksas2024} for the parametrization of the potential. 
Using the Higgs doublet bilinears
\begin{equation}
  P_{ab}
:=
  H_{a}^{\dagger} H_{b}
\enspace , 
\label{eq:higgsdoublet-bilinears}
\end{equation}
gives us
\begin{align}
V(H_{1},H_{2}) 
& =
  \mu_{1} P_{11}
+ \mu_{2} P_{22}
+ \mu_{3} P_{12}
+ \mu_{3}^{\ast} P_{21}
\nonumber \\ & 
+ \lambda_{1} P_{11}^{2}
+ \lambda_{2} P_{22}^{2}
+ \lambda_{3} P_{11} P_{22}
+ \lambda_{4} P_{12} P_{21}
\nonumber \\ & 
+ \lambda_{5} P_{12}^{2}
+ \lambda_{5}^{\ast} P_{21}^{2}
+ P_{11} ( \lambda_{6} P_{12} + \lambda_{6}^{\ast} P_{21} ) 
+ P_{22} ( \lambda_{7} P_{12} + \lambda_{7}^{\ast} P_{21} ) 
\enspace .
\label{eq:higgspot}
\end{align}

As a minimal extension the GNM adds only a single fermionic gauge singlet $N$ to the fermion content of the SM. Being a gauge singlet allows $N$ to have a Majorana mass term, giving the Majorana Lagrangian
\begin{equation}
  \mathcal{L}_{\mathrm{N}}
=
  \sfrac{1}{2} \bar{\hat{N}} i\slashed{\partial} N
- \sfrac{1}{2} M \bar{\hat{N}} P_{R} N
+ h.c.
\enspace ,
\label{eq:neutrino}
\end{equation}
where we use the notation of \cite{pal2011} for the 
Lorentz Covariant Conjugation (LCC) 
\begin{equation}
  \hat{N} = \gamma_{0} C N^{\ast}
\enspace ,
\label{eq:LCC}
\end{equation}
and parametrize the Majorana degrees of freedom by the right-handed components $P_{R} N$. With these fields 
we can now give the Yukawa Lagrangian used in \cite{drauksas2024}
\begin{equation}
- \mathcal{L}_{\mathrm{Y}}
=
  \sum_{a,j,k}
  (Y_{L}^{(a)})_{jk} \bar{\ell}_{Rj} P_{L} (L_{k}.H_{a}^{\ast})
+ \sum_{a,k}
  (Y_{N}^{(a)})_{k} 
  \frac{1}{2} [ \bar{\hat{N}} P_{L} (L_{k} . \epsilon H_{a}) 
               + (\epsilon H_{a} .\bar{\hat{L}}_{k}) P_{R} N ]
+ h.c.
\enspace ,
\label{eq:yukawa}
\end{equation}
where we indicate the $SU(2)$ index contraction with $(a.b)$ and understand
\begin{equation}
  \epsilon H_{a} 
=
  \begin{pmatrix}
    0 & 1
  \\
    -1 & 0
  \end{pmatrix}
  \begin{pmatrix}
    H_{a}^{+} 
  \\
    H_{a}^{0} 
  \end{pmatrix}
=
  \begin{pmatrix}
    H_{a}^{0} 
  \\
  - H_{a}^{+} 
  \end{pmatrix}
\enspace .
\label{eq:adjoint-higgs}
\end{equation}

The total bare Lagrangian sums to 
\begin{equation}
  \mathcal{L}_{\mathrm{GNM}}
=
  \mathcal{L}_{\mathrm{gauge}}
+ \mathcal{L}_{\mathrm{F-G}}
+ \mathcal{L}_{\mathrm{H}}
+ \mathcal{L}_{\mathrm{N}}
+ \mathcal{L}_{\mathrm{Y}}
\enspace .
\label{eq:L-GMN}
\end{equation}

\subsection{The physical particles}

The bare Lagrangian in eq.~(\ref{eq:L-GMN}) is the simplest schematic of the GNM. Using eqs.~\eqref{eq:gauge}, ~\eqref{eq:gaugeferm}, ~\eqref{eq:higgs}, ~\eqref{eq:neutrino},  and~\eqref{eq:yukawa} one can write down the Feynman rules for particle interactions in the flavour basis. However, in this basis, the propagators of the particles are non-diagonal and thus it is not simple to calculate with them. Therefore we go to the mass basis, in which the particles correspond to the observable ones. In the following sections we present the rotation to the mass basis for the gauge bosons, the Higgs bosons, and the leptons. 

\subsubsection{Gauge sector}
\label{sec:counterterm}

 We take the definitions of~Ref.~\cite{degrande2015}, which corresponds to taking $\eta_{Z}=\eta_{\theta} =1$ in Ref.~\cite{romao2012}
\begin{equation}
    \begin{matrix}
        A_\mu=\sin\theta_{\mathrm{W}}W_\mu^3+\cos\theta_{\mathrm{W}}B_\mu \enspace ,
        \\[2pt]
        Z_\mu=\cos\theta_{\mathrm{W}}W_\mu^3-\sin\theta_{\mathrm{W}}B_\mu \enspace .
    \end{matrix}
    \label{eq:zandphoton}
\end{equation}
The relations of the coupling constants in eq.~(\ref{eq:covderiv}) to the electric coupling constant 
\begin{equation}
  e 
= 
  g^{\prime} \cos\theta_{\mathrm{W}} 
= 
  g \sin\theta_{\mathrm{W}}
\label{eq:Weinberg-angle}
\end{equation}
use the same Weinberg angle $\theta_{\mathrm{W}}$. 
For completeness we rotate the fields $W^{1,2}$ into their electro-magnetic charge eigenstates as usual
\begin{equation}
    W_{\mu}^{\mp}=\frac{1}{\sqrt{2}}\left(W_{\mu}^{1}\pm iW_{\mu}^{2}\right)\enspace.
    \label{eq:wpm}
\end{equation}

\subsubsection{Higgs sector}
\label{sec:higgsphysical}
Inserting the Higgs doublets, eq.~\eqref{eq:1sthiggs}, with their vevs into the potential in eq.~\eqref{eq:higgspot} allows to define the Higgs mass matrix as the terms bilinear in the scalar fields. In the Higgs basis the field $h_2^+$ is already in its mass eigenstate. For the neutral fields we diagonalise their mass matrix to get the mass eigenstates. 

Diagonalising this mass matrix with the orthogonal matrix $R$~\cite{haber2006,haber2011,branco2012} one can relate the neutral Higgs fields
\begin{equation}
    \begin{pmatrix}h_{0}\\
H_{0}\\
A
\end{pmatrix}=R^T\begin{pmatrix}h_{1}\\
h_{2}\\
h_{3}
\end{pmatrix}
\enspace ,
\label{eq:higgsmix}
\end{equation}
to the mass eigenstates: $h_{0}$ corresponds to the SM Higgs field. Assuming a $CP$-conserving (CPC) potential in eq.~\eqref{eq:higgspot}, $H_0$ is the second scalar, and $A$ the pseuodoscalar Higgs field. The general matrix $R$~\cite{haber2006,haber2011,branco2012} is
\begin{equation}
    R^T=\begin{pmatrix}c_{12}c_{13} & -s_{12} & -c_{12}s_{13}\\
s_{12}c_{13} & c_{12} & -s_{12}s_{13}\\
s_{13} & 0 & c_{13}
\end{pmatrix}
\enspace ,
\label{eq:rhiggsmix}
\end{equation}
where $c_{ab}$ and $s_{ab}$ are cosine and sine of the mixing angles between the Higgses.
Having the mixing matrix $R$ it is instructive to write the second Higgs doublet in eq.~\eqref{eq:1sthiggs} in terms of the mass eigenstates
\begin{equation}
    \begin{pmatrix}
        h_{2}^{+}\\
        \frac{1}{\sqrt{2}}( h_{2} + i h_{3} ) 
    \end{pmatrix}
    =
    \begin{pmatrix}
        h_{2}^{+}\\
        \frac{1}{\sqrt{2}}\sum_{S}(R_{h_2S} + i R_{h_3S})S
    \end{pmatrix}\,,
\end{equation}
where $S=h_0, H_0, A$. This form of the second doublet implies a natural parametrization by
\begin{equation}
    R_S^\pm=R_{h_2S} \pm i R_{h_3S} 
    \qquad \text{with} \qquad
    R_S^{+} R_S^{-}=1-R^2_{h_1S}\,,
\end{equation}
which we use extensively in later sections.

\subsubsection{Leptons}

We choose a basis in which the charged lepton mass matrix is diagonal. Then the Yukawa coupling between charged leptons and the first Higgs doublet is
\begin{equation}
    (Y_{L}^{(1)})_{jk}\equiv\frac{m_j\sqrt{2}}{v}\delta_{jk}.
    \label{eq:1stlepyukawa}
\end{equation}
The Yukawa couplings $Y_{L}^{(2)}$ between charged leptons and the second Higgs doublet are completely general.
 
 Using the Dirac mass $\frac{v}{\sqrt{2}}\vec{Y}^{\left(1\right)}_N$ with
\begin{equation}
    \vec{Y}^{\left(1\right)}_N=\left((Y^{\left(1\right)}_N)_e,(Y^{\left(1\right)}_N)_\mu,(Y^{\left(1\right)}_N)_\tau\right)^T_{\,}
\end{equation}
one can construct from eqs.~\eqref{eq:neutrino} and~\eqref{eq:yukawa} the neutrino mass matrix
\begin{equation}
    M_\nu=\begin{pmatrix}
        0_{3 \times 3} &\frac{v}{\sqrt{2}}(\vec{Y}^{\left(1\right)}_N)
        \\
        \frac{v}{\sqrt{2}}(\vec{Y}^{\left(1\right)}_N)^T &M
    \end{pmatrix}.
\end{equation}
Using the Takagi factorization~\cite{takagi1924}, we diagonalise it with a unitary matrix $U$
\begin{equation}
    U^{T} M_\nu U =\mathrm{diag}\left(m_1,m_2,m_3,m_4\right).
\end{equation} 
The active neutrino mass $m_{3}$ is generated by the seesaw mechanism, and $m_{4} = |M| + m_{3}$ is typically
associated with the sterile neutrino mass\footnote{For this paper we take normal hierarchy of neutrinos, i.e.
$m_{1}<m_{2}<m_{3}$}. The masses $m_1$ and $m_2$ are equal to $0$ at tree-level. 

The matrix $U$ can be decomposed as~\cite{grimus1989,grimus2001,drauksas2024}
\begin{equation}
    U=\begin{pmatrix}
        U_L \\
        U_R^\ast
    \end{pmatrix}
    \enspace ,
    \label{eq:udecomp}
\end{equation}
where $U_{L}$ is built from the columns $\vec{\,\mathrm{v}}_{k}$ of the $3\times3$ PMNS matrix~\cite{pontecorvo1958,maki1962} 
%$
\begin{equation}
V = \begin{pmatrix}
        \vec{\,\mathrm{v}}_{1} & \vec{\,\mathrm{v}}_{2} & \vec{\,\mathrm{v}}_{3} 
%        {\mathbf{v}}_{1} & {\mathbf{v}}_{2} & {\mathbf{v}}_{3} 
    \end{pmatrix}
\end{equation}
%$ 
as
\begin{equation}
    U_L = \begin{pmatrix}
        \vec{\,\mathrm{v}}_{1}^{*} & \vec{\,\mathrm{v}}_{2}^{*} & c \vec{\,\mathrm{v}}_{3}^{*} & i s \vec{\,\mathrm{v}}_{3}^{*} 
    \end{pmatrix}
    \quad \text{and} \quad 
    U_R^\ast=\begin{pmatrix}
        0 & 0 & is & c
    \end{pmatrix}
    \enspace .
\end{equation}
The sine $s$ and cosine $c$ of the seesaw angle $\theta$  can be expressed as 
\begin{equation}
  s^{2} \equiv \sin^{2}\theta = \frac{m_{3}}{m_{3}+m_{4}}
\;\;\mathrm{and}\;\;
  c^{2} \equiv \cos^{2}\theta = \frac{m_{4}}{m_{3}+m_{4}}. 
\label{eq:seesaw-angle}
\end{equation}

We transform the neutrino fields into mass eigenstates $n$ using this mixing matrix $U$:
\begin{equation}
  \nu_{j}
=
  ( U_L )_{jk} P_{L}n_{k}
\enspace,
\label{eq:acttrsf}
\end{equation}
where the flavour index $j$ goes over the charged lepton flavours,
and 
\begin{equation}
  P_{R}N
=
  ( U_R )_{k}^{\ast}P_{R}n_{k}
\enspace .
\label{eq:stertrsf}
\end{equation}
The sum over the mass eigenstate index $k=1,2,3,4$ is implied.

For the Yukawa couplings we follow Ref.~\cite{dudenas2017}
\begin{equation}
  \vec{Y}_{N}^{(1)}
=
- i \frac{\sqrt{2m_{3}m_{4}}}{v}\vec{\,\mathrm{v}}_{3}
=:
- i y \vec{\,\mathrm{v}}_{3}
\quad\text{and}\quad
  \vec{Y}_{N}^{(2)}
=
  d \vec{\,\mathrm{v}}_{2}
+ d' \vec{\,\mathrm{v}}_{3}
\enspace ,
\label{eq:yukawacouplings}
\end{equation}
where $d>0\in\mathbb{R}$ and $d'\in\mathbb{C}$. The first Yukawa coupling generates the seesaw mass term at tree-level. The second Yukawa coupling is responsible for the mass of $n_2$ at $1$-loop-level~\cite{grimus1989}. 

In this paper we take the ''measured'' masses 
\begin{equation}
  m_2^{\text{pole}}
=
  \sqrt{\Delta m_{21}^2}
\;\;\mathrm{and}\;\;
  m_3^{\text{pole}}
=
  \sqrt{\left|\Delta m_{32}^2\right|+\Delta m_{21}^2},
\end{equation}
with the values found in Ref.~\cite{navas2024}
\begin{equation}
    \Delta m_{21}^2=(7.53\pm0.18)\cdot 10^{-5}\;\mathrm{eV}^2\;\;\mathrm{and}\;\;\left|\Delta m_{32}^2\right|=(2.455\pm0.028)\cdot 10^{-3}\;\mathrm{eV}^2.
\end{equation}
For the PMNS matrix $V$
\begin{equation}
    V=\begin{pmatrix}
    c_{12}c_{13}&s_{12}c_{13}&s_{13}e^{-i\delta}
        \\
        -s_{12}c_{23}-c_{12}s_{23}s_{13}e^{i\delta} &  c_{12}c_{23}-s_{12}s_{23}s_{13}e^{i\delta}&s_{23}c_{13}
        \\
          s_{12}s_{23}-c_{12}c_{23}s_{13}e^{i\delta} &  -c_{12}s_{23}-s_{12}c_{23}s_{13}e^{i\delta}&c_{23}c_{13}
    \end{pmatrix},
\end{equation}
where $s_{ij}=\sin{\theta_{ij}}$ and $c_{ij}=\sqrt{1-s_{ij}^2}$, we take the values from Ref.~\cite{navas2024}\footnote{These sines and cosines are not the same as in eq.~\eqref{eq:rhiggsmix} and will only be used in this section for the PMNS matrix.}
\begin{equation}
    \theta_{12}=35.74^{\degree}, \;\;\theta_{23}=51.9^{\degree}\;\;\theta_{13}=8.89^{\degree}\;\;\mathrm{and}\;\;\delta=197^{\degree}.
\end{equation}

\section{The Higgs decay rates}
\label{sec:Higgsdecays}
\begin{figure}[hbt!]
\centering
\includegraphics{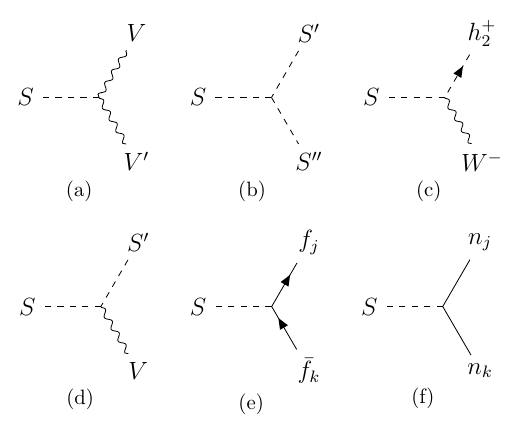}
% \subfigure[]{ 
%     \feynmandiagram 
%     [inline=(d.base), small, horizontal=d to b] {
% a [particle=\(V\)]--[boson]b --  [boson]c[particle=\(V'\)],
% b -- [scalar] d [particle=\(S\)],
% };
% }
% \subfigure[]{ 
%     \feynmandiagram 
%     [inline=(d.base), small, horizontal=d to b] {
%         a [particle=\(S'\)]--[scalar]b --  [scalar]c[particle=\(S''\)],
%         b -- [scalar] d [particle=\(S\)],
%     };
% }
% \subfigure[]{ 
%     \feynmandiagram 
%     [inline=(d.base), small, horizontal=d to b] {
%         a [particle=\(h_2^+\)]--[anti charged scalar]b --  [boson]c[particle=\(W^-\)],
%         b -- [scalar] d [particle=\(S\)],
%     };
% }
% \\
% \subfigure[]{ 
%     \feynmandiagram 
%     [ inline=(d.base), small, horizontal=d to b] {
%         a [particle=\(S'\)]--[scalar]b --  [boson]c[particle=\(V\)],
%         b -- [scalar] d [particle=\(S\)],
%     };
% }
% \subfigure[]{ 
%     \feynmandiagram 
%     [inline=(d.base), small, horizontal=d to b] {
%         a [particle=\(f_j\)]--[anti fermion] b --[anti fermion]c[particle=\(\bar{f}_k\)],
%         b -- [scalar] d [particle=\(S\)],
%     };
% }
% \subfigure[]{ 
%     \feynmandiagram 
%     [inline=(d.base), small, horizontal=d to b] {
%         a [particle=\(n_j\)]--b --  c[particle=\(n_k\)],
%         b -- [scalar] d [particle=\(S\)],
%     };
% }
\caption{The amplitude of a neutral Higgs boson $S=h_0,H_0,A$ going into: (a) to gauge bosons $V,V'=\gamma,Z,W$; (b) to neutral Higgses $S',S''=h_0,H_0,A$; (c) a charged Higgs boson and a $W$ boson; (d) to another neutral Higgs boson and a neutral gauge boson; (e) two charged fermions $f_j$ and $\bar{f}_k$; (f) two Majorana neutrinos $n_j$ and $n_k$.}
\label{fig:feyndiaghiggs}
\end{figure}

In this section we present all two particle Higgs decays. The Feynman diagrams are pictured in Fig.~\ref{fig:feyndiaghiggs}. Using the GNM Lagrangian in eq.~\eqref{eq:L-GMN} we express the amplitudes. Summing over spins or polarizations of the final state particles and using the phase space element in appendix~\ref{sec:phasespace}, we write down the decay rates. Where applicable, we compare the results with the literature.

\subsection{Higgs decay into gauge bosons}

Here we present the Higgs boson decays into gauge bosons. We construct the amplitude using eqs.~\eqref{eq:covderiv},~\eqref{eq:higgs},~\eqref{eq:zandphoton}, ~\eqref{eq:wpm}, and 
Fig.~\ref{fig:feyndiaghiggs}. This amplitude is
\begin{equation}
     i\mathcal{M}_{S\to VV'}=\frac{1}{2}ig^{2}_{VV'}vR_{h_{1}S}g^{\mu\nu}\epsilon^{\ast}{}^{\mu}\left(p_{1}\right)\epsilon^{\ast}{}^{\nu}\left(p_{2}\right),
\end{equation}
where $S=h_0,H_0,A$ corresponds to the Higgs mass eigenstates, $V,V'=\gamma,W,Z$, and 
\begin{equation}
    g_{VV'}=
    \begin{cases}
        0,&\mathrm{if}\;\;V\;\mathrm{or}\;V'=\gamma \\
%        0,&\mathrm{if}\;\;V=V'=\gamma\;\;\mathrm{or}\;\;V=\gamma\;\;\mathrm{and}\;\;V'=Z \\
        g,&\mathrm{if}\;\;V=V'=W  \\
        \frac{g}{\cos{\theta_\mathrm{W}}},&\mathrm{if}\;\;V=V'=Z  \\
    \end{cases}\enspace .
\end{equation}
The momenta of the outgoing particles are $p_i$ with $i=1,2$.

To get the decay rate, we square this amplitude and sum over polarizations of the final state particles. Using the result for the phase space element in Table~\ref{table_2pLips} in appendix~\ref{sec:phasespace} we write down the decay rate
\begin{equation}
    \Gamma_{S\to VV'}=\frac{g_{VV'}^{2} m_{S}^3R_{h_{1}S}^{2}}{64\pi m_{V}^{2}}\lambda'_{SV}f_V^{},
\end{equation}
where 
\begin{equation}
    \lambda'_{SV}=\sqrt{1-4x_{SV}^{}}\left(1-4x_{SV}^{}+12x^{2}_{SV}\right),
\end{equation}
with $x^{}_{SV}=\frac{m_V^2}{m_{S}^2}$, and
\begin{equation}
    f_V^{}=
    \begin{cases}
        \frac{1}{2},&\mathrm{if}\;\;V=Z%,\gamma
        \\
        1,&\mathrm{if}\;\;V=W
    \end{cases}\enspace,
\end{equation}
is the statistical factor. %We took $V=V'$ for these parameters.
% We omit the $S=h_0$ case since its mass is too small. 

For CPC, i.e. $s_{13}=0$\footnote{Remember, these are the sines coming from the Higgs mixing matrix in eq.~\eqref{eq:rhiggsmix}.}, and using $v=\frac{2m_W}{g}$, the decay rates match the ones found in Refs.~\cite{gunion2018,branco2012}. For the decay $h_{0} \to WW$ of Ref~\cite{branco2012} we have to identify their $\sin^{2}{(\alpha-\beta)}$ with our $R_{h_{1}S}^{2}$. Taking  our  $R_{h_1S}^2$ and identifying it with $\cos^2{(\beta-\alpha)}$ we get the decay rates that are equivalent to the ones in Ref.~\cite{gunion2018}. In the case of $CP$ conservation, the pseudoscalar $A$  does not decay into two gauge bosons.

\subsection{Higgs decay into lighter Higgses}
\label{sec:scalartoscalar}
The amplitudes for this decay are
\begin{equation}
    i\mathcal{M}_{S\to S'S''}=-iv\Lambda_{SS'S''},
    \label{eq:sss}
\end{equation}
where 
\begin{align}
  \Lambda_{SS'S''}
&=
  \frac{1}{v^{2}} R_{h_{1}S} 
  [ ( \delta_{S'S''} - R_{h_{1}S'}R_{h_{1}S''} )
  ( \lambda_{3} v^{2} 
  - 2 m_{h_{2}^{+}}^{2} 
  + m_{S'}^{2} 
  )
  + \delta_{S'S''} m_{S'}^{2} 
  ]
\nonumber\\&
+ \mathrm{Re}[\lambda_{7} R_{S}^{+}R_{S'}^{+}R_{S''}^{-} ]
+ \text{cyclic permutations of $\{S,S',S''\}$ }
\enspace .
\label{eq:lsss}
\end{align}
For this coupling we used the expressions for $\lambda_1$, $\lambda_4$, $\lambda_5$ and $\lambda_6$ from appendix~\ref{sec:lambdasection}. Using the result in appendix~\ref{sec:phasespace}, eq.~\eqref{eq:dplipsfinal} we write the decay rate
\begin{equation}
    \Gamma_{S\to S'S''}=\frac{v^2\Lambda_{SS'S''}^2}{16\pi m_{S}^3}\left(1-\frac{\delta_{S'S''}}{2}\right)\sqrt{\left(m_{S}^{2}-m_{S'}^{2}-m_{S''}^{2}\right)^{2}-4m_{S'}^{2}m_{S''}^{2}},
\end{equation}
where $S\neq S'S''$ and the brackets with the Kronecker delta account for the statistical factor.

In CPC all $\lambda$s are real, and $\lambda_7=0$, and the mixing between the pseudoscalar $A$ and the real scalars $h_{0}$ and $H_{0}$ vanishes, giving $R_{h_{1,2}A}=R_{h_3h_0}=R_{h_3H_0}=0$. Then $\Lambda_{AS'S''}$ vanishes and the pseudoscalar does not decay into the other Higgs bosons. For the scalar $H_0$, since it cannot decay into $A$, it leaves us with the coupling $\Lambda_{H_0h_0h_0}$, which is
\begin{align}
    \Lambda_{H_0h_0h_0}^{\mathrm{CPC}} 
&=
  \frac{R_{h_{1}H_0}}{v^{2}} 
  [ ( 1 - 3 R_{h_{1}h_0}^2 )
    ( \lambda_{3} v^{2} 
    - 2 m_{h_{2}^{+}}^{2} 
    )
  + ( 1 - R_{h_{1}h_0}^2 )
    2 m_{h_0}^{2} 
  - R_{h_{1}h_0}^2 
    m_{H_0}^{2} 
  ]
\nonumber\\&
+ \mathrm{Re}[\lambda_{7} R_{h_0}^{+}
  ( 2 R_{H_0}^{+}R_{h_0}^{-} + R_{h_0}^{+}R_{H_0}^{-} ) ]
\enspace .
\end{align}

The GNM and IDM both have a similar coupling structure, but Higgs doublets do not mix in the IDM. To take the IDM limit for the GNM, in addition to the previous parameter choices for CPC, we also take $s_{12}=1$. Then the elements of matrix $R$ in eq.~\eqref{eq:rhiggsmix} become
\begin{equation}
\label{R-IDM-limit}
    R=
    \begin{pmatrix}
        R_{h_1h_0} & R_{h_2h_0} & R_{h_3h_0}
        \\
        R_{h_1H_0} & R_{h_2H_0} & R_{h_3H_0}
        \\
        R_{h_1A} & R_{h_2A} & R_{h_3A}
    \end{pmatrix}
    \to
    \begin{pmatrix}
        0 & 1 & 0
        \\
        -1 & 0 & 0
        \\ 
        0 & 0 & 1
    \end{pmatrix},
\end{equation}
with 
\begin{equation}
    R^\pm_{S}=R_{h_2S}\pm iR_{h_3S}\to-\delta_{S h_0}\pm i\delta_{SA}.
\end{equation}
The coupling in the IDM limit is
\begin{equation}
  \Lambda^{\mathrm{IDM}}_{H_0h_0h_0}
=
  \lambda_3+\frac{2m_{h_0}^2-2m_{h_2^+}^2}{v^2}
=
  \lambda_3+\lambda_4+2\mathrm{Re}\left[\lambda_5\right]
=: 
  \lambda_{345}
\label{eq:l345}
\enspace , 
\end{equation}
where the transformation laws in appendix~\ref{sec:lambdasection} give the second equality. 
In Refs.~\cite{branco2012,krawczyk2016} the factor of $2$ in front of $\lambda_5$ in eq.~\eqref{eq:l345} does not appear due to differing conventions in the potential in eq.~\eqref{eq:higgspot}.

In this limit the decay rate is
\begin{equation}
  \Gamma_{H_0\to h_0h_0}^\mathrm{IDM\;limit}
=
  \frac{v^2\left(\lambda_{345}\right)^2}{32\pi m_{H_0}}
  \sqrt{1 - \frac{4m^2_{h_0}}{m_{H_0}^2}}
\enspace ,
\label{eq:decayinert}
\end{equation}
which is equivalent to the one found in Ref.~\cite{krawczyk2016} (pg. $188$),  except in the reference, the decay is with particles swapped, i.e. $h_0\to H_0H_0$.

\subsection{Higgs decays into a charged Higgs and \texorpdfstring{$W$}{W} boson}

If the neutral Higgs bosons are more massive than the charged Higgs boson, they can decay into it and a charged gauge boson. Using eq.~\eqref{eq:higgs} we read the amplitude for such a decay
\begin{equation}
    i\mathcal{M}_{S\to W^-h_2^+}=i\frac{g}{2}R^+_S\left(p_1^\mu+q^\mu\right)\epsilon^{*\mu}\left(p_2^\mu\right).
\end{equation}
Using the result of eq.~\eqref{eq:dplipsfinal} in appendix~\ref{sec:phasespace} for the general phase space element, we write down the decay rate as
\begin{equation}
    \Gamma_{S\to W^- h_2^+}^{}
= \frac{g^2}{64 \pi  m_W^2m_S^3} 
  R_{S}^{+} R_{S}^{-}
  \left(\left(m_S^2-m_W^2-m_{h_2^+}^2\right)^2-4m_W^2m_{h_2^+}^2\right)^{3/2}.
\end{equation}

In a $CP$ conserving case, the pseudoscalar decouples and cannot decay into these bosons. The $H_0$ decay then becomes equivalent to the one found in~\cite{gunion2018}. 

\subsection{Higgs decays into a neutral Higgs and \texorpdfstring{$Z$}{Z} boson}

If one of the neutral Higgs bosons is more massive than the other neutral Higgses, it can decay into it and a $Z$ boson. Using eq.~\eqref{eq:higgs} we read the amplitude for this process
\begin{equation}
    i\mathcal{M}_{S\to S'Z}=
    \frac{g}{2\cos{\theta_\mathrm{W}}} 
   \mathrm{Im}\left[R_S^+R_{S'}^-\right]
    \left(p_1^{\mu }+q^{\mu }\right)  
    \epsilon^{*\mu},
\end{equation}
As we did in the previous section, we use the general phase space element, eq.~\eqref{eq:dplipsfinal}, to write down the decay rate
\begin{equation}
    \Gamma_{S\to S'Z}=
    \frac{g^2}{64\pi m_S^3m_Z^2\cos^2\theta_{\mathrm{W}}}
    \left(\mathrm{Im}\left[R_S^+R_{S'}^-\right]\right)^2
    \left(\left(m_S^2-
    m_{S'}^2-m_Z^2\right)^2-
    4m_Z^2m_{S'}^2\right)^{3/2},
\end{equation}
where $S\neq S'$. This decay rate is equivalent to those found in Refs.~\cite{gunion2018,branco2012}.

\subsection{Higgs decays into charged fermions}

Here we present Higgs decays into charged fermions. For these decays we take the fermions\footnote{Our calculations do not include top quarks, as they are too heavy.} in the final state phase space to be massless, since their masses are much smaller than the mass of the decaying Higgs.

For the decay (e) of Fig.~\ref{fig:feyndiaghiggs} the sum of the Yukawa couplings, eq.~\eqref{eq:yukawa}, 
\begin{align}
   ( \mathcal{A}_{F} )_{jk}
&=
  R_{h_{1}S} (Y_{F}^{(1)})_{jk}
+ R_{S}^{-} (Y_{F}^{(2)})_{jk}
\enspace , 
\end{align}
gives the amplitude 
\begin{align}
   i\mathcal{M}_{S\to\ell_j\ell_k}
&=
  \frac{i}{\sqrt{2}}
  \bar{u}_{j} \left[ 
    ( \mathcal{A}_{F} )_{jk} P_{L}
  + ( \mathcal{A}_{F}^{\dagger} )_{jk} P_{R}
  \right] v_{k}
\enspace .
\end{align}
The decay rate is then 
\begin{align}
  \Gamma_{S\to\ell_j\ell_k}
&=
  \frac{m_{S}}{32\pi}
  \left( | (\mathcal{A}_{F} )_{jk} |^{2}
    + | (\mathcal{A}_{F} )_{kj} |^{2}
%  + (j \leftrightarrow k )
  \right)
\nonumber\\[0pt] 
&=
  \frac{m_{S}}{32\pi}\bigg(
    \delta_{jk} R_{h_{1}S}^{2} \frac{4 m_{j}^{2}}{v^{2}}
  + \delta_{jk} R_{h_{1}S} \frac{4 \sqrt{2} m_{j}}{v} 
    \mathrm{Re}[ R_{S}^{-} (Y_{F}^{(2)})_{jk} ]
\nonumber\\[0pt] & \qquad
  + R_{S}^{+} R_{S}^{-} 
    \left( | (Y_{F}^{(2)})_{jk} |^{2} + | (Y_{F}^{(2)})_{kj} |^{2} 
    \right)
\bigg)
\enspace .
    \label{eq:higgsintoleptons}
\end{align}
The first term comes from the first Higgs doublet, while the others come because of the second Higgs doublet.

Taking only the term coming from the first Higgs doublet, setting $S=h_0$ and $R_{h_1h_0}=1$, and using eq.~\eqref{eq:1stlepyukawa} we get that the decay rate simplifies to the SM Higgs decay rate
\begin{equation}
    \Gamma^\mathrm{SM}_{h_0\to\ell_j\ell_j}=\frac{m_{h_0 }m_j^2}{8\pi v^2}
\enspace ,
\end{equation}
which can be found in the literature, e.g.~\cite{gunion2018}.

For non-diagonal lepton Yukawa couplings $Y_{F=L}^{(2)}$ the decays, eq.~(\ref{eq:higgsintoleptons}), also include charged lepton flavour violating (cLFV) currents.

\subsection{Higgs decays into two neutrinos}

%In this subsection we calculate how the Higgs bosons
%decay into neutrinos. 
For the kinematics we take the active neutrinos to be massless in the phase space of the final state.
The sum of the Yukawa couplings eq.~\eqref{eq:yukawa} 
\begin{align}
   ( \mathcal{A}_{N} )_{jk}
&=
R_{h_{1}S} (
  U_{L}^{T} \cdot 
  Y_{N}^{(1)}
  \cdot U_{R}^{\ast} 
)_{jk}
+
R_{S}^{+} (
  U_{L}^{T} \cdot 
  Y_{N}^{(2)}
  \cdot U_{R}^{\ast} 
  )_{jk}
\enspace , 
\end{align}
has to be symmetrized and gives the general amplitude for these decays 
\begin{align}
  i\mathcal{M}_{S\to n_j n_k}
=&
- \frac{i}{\sqrt{2}}
  \bar{u}_{j} \left[ 
    ( \mathcal{A}_{N} )_{jk} P_{L}
  + ( \mathcal{A}_{N} )_{kj} P_{L}
  + ( \mathcal{A}_{N}^{\dagger} )_{jk} P_{R}
  + ( \mathcal{A}_{N}^{\dagger} )_{kj} P_{R}
  \right] v_{k}
\enspace .
\end{align}
With the splitting of $U$ into $U_{L}$ and $U_{R}$, the amplitude $\mathcal{A}_{N}$ also splits into a direct product of the two vectors $(U_{L}^{T} \cdot Y_{N}^{})$ and $U_{R}$. We can easily see
\begin{align}
  U_{L}^{T} \cdot \vec{\,\mathrm{v}}_{j}
=
\begin{pmatrix}
   \vec{\,\mathrm{v}}_{1}^{*} & \vec{\,\mathrm{v}}_{2}^{*} & c \vec{\,\mathrm{v}}_{3}^{*} & i s \vec{\,\mathrm{v}}_{3}^{*} 
\end{pmatrix}^{T}
\cdot
\vec{\,\mathrm{v}}_{j}
= 
  ( \begin{array}{cccc} 
    \delta_{j1} & \delta_{j2} & c \delta_{j3} & is \delta_{j3} 
    \end{array} )^{T}
\enspace ,
\end{align}
which gives the simplifying relations
\begin{align}
  U_{L}^{T} \cdot \vec{Y}_{N}^{(1)}
=
  U_{L}^{T} \cdot ( - i y \vec{\,\mathrm{v}}_{3} )
= 
- i y
  ( \begin{array}{cccc} 0 & 0 & c & is \end{array} )^{T}
\enspace ,
\end{align}
and
\begin{align}
  U_{L}^{T} \cdot \vec{Y}_{N}^{(2)}
=
  U_{L}^{T} \cdot ( d \vec{\,\mathrm{v}}_{2} + d' \vec{\,\mathrm{v}}_{3} )
= 
  ( \begin{array}{cccc} 0 & d & c d' & is d' \end{array} )^{T}
\enspace .
\end{align}
With the abbreviations 
\begin{align}
B_{S}^{+} = y R_{h_{1}S} + i d' R_{S}^{+}
\quad\text{and}\quad
B_{S}^{-} = y R_{h_{1}S} - i d'^{\ast} R_{S}^{-}
\end{align}
we can write $\mathcal{A}_{N}$ in matrix form as
\begin{align}
  ( \mathcal{A}_{N} )_{jk}
&=
  ( \begin{array}{cccc} 0 
  & d R_{S}^{+} 
  & - i c B_{S}^{+} 
  & s B_{S}^{+} \end{array} )^{T}_{j}
  \begin{pmatrix}
        0 & 0 & is & c
  \end{pmatrix}_{k}
\nonumber\\[0pt] 
&=
  [ d R_{S}^{+} \delta_{j2} + B_{S}^{+} ( s \delta_{j4} - i c \delta_{j3} ) ]
  ( i s \delta_{k3} + c \delta_{k4} )
\enspace .
\end{align}
The symmetrization in $(jk)$ then gives the index structure
\begin{align}
\label{eq:2ANjk}
  2 ( \mathcal{A}_{N} )_{(jk)}
&=
  d R_{S}^{+} ( i s \Delta^{23}_{jk} + c \Delta^{24}_{jk} )
%\nonumber\\[0pt] &
+ B_{S}^{+} 
  ( 2 s c ( \delta_{j3} \delta_{k3} + \delta_{j4} \delta_{k4} )
  - i ( c^{2} - s^{2} ) \Delta^{34}_{jk} 
  )
\enspace ,
\end{align}
where 
\begin{equation}
  \Delta^{\ell m}_{jk}
=
  \delta_{j\ell}\delta_{km}+\delta_{jm}\delta_{k\ell}
\enspace .
\end{equation}
From this amplitude we see that the Higgs bosons will decay into these five final states: $n_2n_3$, $n_2n_4$, $n_3n_3$, $n_3n_4$, and $n_4n_4$. These decay rates are:
\begin{equation}
  \Gamma_{S\to n_2n_3}
=
  \frac{1}{2 m_{S}} % Flux
  \frac{1}{8 \pi}  % Phasespace
  \frac{1}{2}  % amplitude 1/sqrt{2} factor
  m_{S}^{2}  % amplitude from spinors (p1.p2)
  2 |2 ( \mathcal{A}_{N} )_{(23)}|^{2} % amplitude from (68)
=
  \frac{m_S}{16 \pi} s^{2} d^{2} R_{S}^{+} R_{S}^{-} 
\enspace ,
\label{eq:n2n3}
\end{equation}
\begin{equation}
  \Gamma_{S\to n_2 n_4}
=
  \frac{(m_S^{2}-m_{4}^{2})^{2}}{32 \pi m_{S}^{3}} 2 |2 ( \mathcal{A}_{N} )_{(24)}|^{2}
=
  \frac{(m_S^{2}-m_{4}^{2})^{2}}{16 \pi m_{S}^{3}} c^{2} d^{2} R_{S}^{+} R_{S}^{-} 
\enspace ,
\end{equation}
\begin{equation}
  \Gamma_{S\to n_3 n_3}
=
  \frac{m_{S}}{32 \pi} |2 ( \mathcal{A}_{N} )_{(33)}|^{2}
=
  \frac{m_{S}}{32 \pi} 4 s^{2} c^{2} | B_{S}^{+} |^{2} 
\enspace ,
\label{eq:n3n3}
\end{equation}
\begin{equation}
  \Gamma_{S\to n_3 n_4}
=
  \frac{(m_S^{2}-m_{4}^{2})^{2}}{32 \pi m_{S}^{3}} 2 |2 ( \mathcal{A}_{N} )_{(34)}|^{2}
=
  \frac{(m_S^{2}-m_{4}^{2})^{2}}{16 \pi m_{S}^{3}} ( c^{2} - s^{2} )^{2} | B_{S}^{+} |^{2} 
\enspace ,
\end{equation}
\begin{align}
  \Gamma_{S\to n_4 n_4}
&=
  \frac{\sqrt{m_S^{2} - 4 m_{4}^{2} }}{64 \pi m_{S}^{2}} 
  \left[
  ( m_S^{2} - 2 m_{4}^{2} ) 2 |2 ( \mathcal{A}_{N} )_{(44)}|^{2}
  - 2 m_{4}^{2}  
  [ 2 ( \mathcal{A}_{N} )_{(44)}^{2}
  + 2 ( \mathcal{A}^{\ast}_{N} )_{(44)}^{2} ]
  \right]
\nonumber\\[0pt] 
&=
  \frac{\sqrt{m_S^{2} - 4 m_{4}^{2} }}{8 \pi m_{S}^{2}} s^{2} c^{2}
  \left[
  ( m_S^{2} - 2 m_{4}^{2} ) |B_{S}^{+}|^{2}
  - 2 m_{4}^{2} \mathrm{Re}[B_{S}^{+ \, 2}]
  \right]
\enspace ,
\end{align}
with 
\begin{align}
  | B_{S}^{+} |^{2} 
= 
  B_{S}^{+} B_{S}^{-} 
&=
  y^2 R_{h_{1}S}^{2} -2 y R_{h_{1}S}  \mathrm{Im}[d'R_S^+] 
+ |d'|^{2} R_{S}^{+} R_{S}^{-}
\enspace ,
\end{align}
and
\begin{align}
  \mathrm{Re}[ B_{S}^{+ \, 2} ]
&=
 \mathrm{Re}[ y^2 R_{h_{1}S}^{2} + 2 i y R_{h_{1}S} d' R_{S}^{+}
  - d'^{2} (R_{S}^{+} )^{2} ]
= |B_{S}^{+}|^2
- 2 ( \mathrm{Re}[ d' R_{S}^{+} ] )^{2}
\enspace .
\end{align}

\section{Scalar dark matter candidates in the GNM}
\label{sec:GNDM}
The IDM is treated in the literature~\cite{ma2006,honorez2007,ilnicka2016,belyaev2018} as one of the simplest extensions of the SM that can provide a DM candidate. The stability of this candidate hinges on an unbroken symmetry, usually a $Z_2$, but sometimes also a $U(1)_{\text{Peccei Quinn}}$~\cite{alves2016}, that prevents the decay. Ref.~\cite{ma2006} connects the IDM to neutrino physics and presents the dependence of the neutrino masses on the parameter $\lambda_{5}$ of the Higgs potential, eq.~(\ref{eq:higgspot}). Ref.~\cite{arina2009} explains the smallness of $\lambda_{5}$ as technical naturalness, as the vanishing of $\lambda_{5}$ enhances the symmetry of the model.  Ref.~\cite{dudenas2022} extends this concept to define the tiny seesaw regime of the GNM, where not only $\lambda_{5}$, but also the $Z_2$ violating neutrino Yukawa couplings are taken to be technically natural, i.e. very small, since their absence enhances the symmetry of the model. 
In this context we ask whether the pseudoscalar $A$ of the GNM behaves like the DM candidate of the IDM in the limit where both models appear most similar. 

Using eq.~(\ref{R-IDM-limit}) and $\lambda_{6,7} \to 0$ simplifies
\begin{equation}
    R_S^+R_S^-  \to  1-\delta_{Sh_0}=\delta_{SH_0}+\delta_{SA}
\enspace ,
\quad
\text{and}
\quad
    B_S^{+} \to  ( y - i d ) \delta_{Sh_0} - d'\delta_{SA} 
\enspace , 
\end{equation}
giving the decay rate to the light neutrinos as
\begin{equation}
  \Gamma_{A\to2\nu}^{\mathrm{IDM~limit}}
=
  \frac{m_A s^2 ( d^2 + 2 c^2 |d'|^2 )}{16\pi}
\enspace .
\label{eq:activedecay}
\end{equation}

The requirement of the GNM to give one neutrino a seesaw mass and another a radiative mass breaks the protective symmetries of the IDM: assigning $N$ to be odd under $Z_{2}$ forbids $Y_{N}^{(1)}$, which is needed for the seesaw. For the radiative mass we need the self-energy function 
\begin{equation}
    \varLambda
= 
  \frac{m_{4}}{32\pi^{2}}
\left[
B_{0}(0,m_4^2,m_A^2)
-
B_{0}(0,m_4^2,m_{H_0}^2)
\right]
\enspace , 
\label{eq:lambdafunction}
\end{equation}
which is proportional to the radiative mass and hence required to be non-zero. 
It is directly proportional to $m_{A}^{2}-m_{H_0}^{2} \propto \lambda_{5}$ which breaks the Peccei Quinn symmetry that the Higgs potential has in the case of $\lambda_{5} \to 0$. 
As a result we have Yukawa couplings in the GNM that break $Z_2$ and depend on $\lambda_{5} \neq 0$. 
An exact parametrization of these couplings, depending on the free parameters of the model, useful in the context of Lepton Flavour violation, can be found in~\cite{dudenas2022}. Using only $d$ and $|d'|$ is enough for our case. Obviously $|d'| \ge 0$, and 
\begin{equation}
  d^{2} 
= 
  \frac{m_{2}^{\text{pole}} m_{3}^{\text{pole}}}{m_{3} |\varLambda|} 
\enspace ,
\end{equation}
which can be seen by taking the determinant of eq.~(2.19) of \cite{dudenas2022}, 
where $m_{3}$ denotes the seesaw mass, that comes from the coupling relation in eq.~(\ref{eq:yukawacouplings}). From eq.~(\ref{eq:seesaw-angle}) we see that $s^{2} \approx \frac{m_3}{m_4}$, which gives the simplified lower limit for the decay rate
\begin{equation}
  \Gamma_{A\to2\nu}^{\mathrm{IDM~limit}}
\ge 
  \frac{m_A s^2 d^2}{16\pi}
= 
  \frac{m_A m_{2}^{\text{pole}} m_{3}^{\text{pole}}}{16\pi m_4 |\varLambda|} 
=: 
  \Gamma^{\mathrm{limit}}_{(m_4)}
\enspace .
\label{eq:activedecay-limit}
\end{equation}
So an upper bound on $|\varLambda|$ in the considered parameter range of $m_4$, $m_A$, and $m_{H_0}$ gives us an upper bound on the lifetime of our IDM-like dark matter candidate. 

As a function $|\varLambda|$ is linear in $m_{H_0}$ and grows the most when $m_{A}$ gets as small as possible and $m_4$ as large as possible. Taking the extreme values $m_4 = 10~\mathrm{TeV}$, $m_A = 60~\mathrm{GeV}$, and $m_{H_0} = 800~\mathrm{GeV}$, we get an upper limit of the lifetime of our IDM-like dark matter candidate of $13$~seconds, which excludes it as a real dark matter candidate. 

\begin{figure}[h]
    \centering
    {\includegraphics{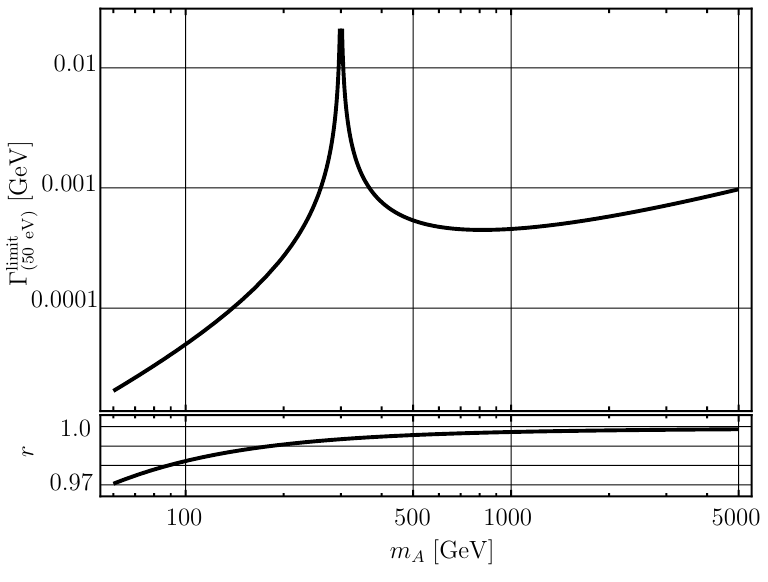}}
    \caption{The lower limit on the decay rate of the pseudoscalar $A$ in the IDM limit. The curve in the upper part shows the limit for a reference value $m_{H_0}=300~\mathrm{GeV}$ and the lower limit of the tiny seesaw of $m_4=50~\mathrm{eV}$. The lower part shows the ratio $r=(\Gamma^{\mathrm{limit}}_{(50~\mathrm{eV})}/\Gamma^{\mathrm{limit}}_{(10~\mathrm{GeV})})/(4 \times 10^{16})$, decreased by the factor $4 \times 10^{16} = (10~\mathrm{GeV})^{2}/(50~\mathrm{eV})^{2}$ for better visibility of the scale. 
    The lifetimes for $m_A=60~\mathrm{GeV}$, corresponding to the two cases are $0.0012~\mathrm{s}$ for $m_4=10~\mathrm{GeV}$ and $3.2\times 10^{-20}~\mathrm{s}$ for $m_4=50~\mathrm{eV}$, respectively.}
    \label{fig:activedecay}
\end{figure}

As a sanity check we present the dependence of our decay rate bound on $m_{A}$, taking a more conservative value for $m_{H_0} = 300~\mathrm{GeV}$ and the upper and lower limit of the tiny seesaw scale with $m_4 = 10~\mathrm{GeV}$ and $m_4 = 50~\mathrm{eV}$ in Fig.~\ref{fig:activedecay}. 

\section{Conclusions}
\label{sec:Conclusions}
In this paper we have calculated the contributions of tree-level processes with two particles in the final state to the decay rates of the Grimus--Neufeld model scalar bosons. The decays into gauge bosons and other bosons match the ones found in the literature, for example, Refs.~\cite{gunion2018,branco2012}. The general form of the decay amplitude into Higgses, eqs.~\eqref{eq:sss} and~\eqref{eq:lsss}, seems to be new, but all special cases that we could find in the literature also match, as an example see~\cite{krawczyk2016}. 
In principle, this general decay amplitude can allow the determination of the free parameters $\lambda_3$ and $\lambda_7$ of the Higgs potential, if the corresponding decays can be measured. 

The decay rate formula into charged leptons, eq.~\eqref{eq:higgsintoleptons}, shows that lepton flavour violating decays are directly proportional to the entries in the second Higgs doublet's Yukawa coupling and that no cancellation of different entries can occur. 

The decay amplitude to neutrinos is more complicated due to their Majorana nature. The simple structure in eq.~(\ref{eq:2ANjk}) shows that the lightest massless neutrino state does not couple to the Higgs bosons and stresses the importance of second Higgs doublet couplings to the active neutrinos. These couplings give a lower bound on the decay rates of the GNM scalars, see eqs.~(\ref{eq:n2n3}), (\ref{eq:n3n3}), and (\ref{eq:activedecay-limit}). From eq.~(\ref{eq:activedecay-limit}) one sees the proportionality of this lower bound to the masses of the light neutrinos and the breaking of the $U(1)_{\text{PQ}}$, as discussed below eq.~(\ref{eq:lambdafunction}).

The IDM or the scoto-seesaw model~\cite{rojas2019} have a very similar coupling structure to our GNM, especially with respect to Lepton Flavour violating observables, see~\cite{dudenas2022}. Both the IDM and the scoto-seesaw model have a scalar DM candidate. 
But the lower bound for the decay rate excludes the GNM scalar dark matter candidate. In this aspect the slight symmetry breaking required in the GNM is enough to have a large phenomenological impact.

%\clearpage{}

\section*{Acknowledgments}
This research has been carried out in the framework of the agreement of Vilnius University with the Lithuanian Research Council No. VS-13.

\appendix
\section{Calculating phase space elements}
\label{sec:phasespace}

% \subsection{Two particles in the final state}
\label{sec:twoparticledlips}
We derive the general result in the reference frame of the decaying particle. The general two particle phase space element is
\begin{equation}
    d\varPi_\mathrm{LIPS}=\frac{1}{\left(2\pi\right)^{6}}\frac{1}{4E_1E_2}d^{3}p_{1}d^{3}p_{2}\left(2\pi\right)^{4}\delta^{4}\left(q-p_{1}-p_{2}\right),
\end{equation}
where $p_{1}$ and $p_{2}$ are the momenta of the outgoing particles and
$q$ is the momentum of the decaying particle. The $\delta$-function enforces energy-momentum conservation
\begin{equation}
    q^\mu=p_1^\mu+p_2^\mu
\enspace ,
\end{equation}
with $\vec{q}=0$ in the decaying particle's rest frame. The energies are defined as
\begin{equation}
    E_i=\sqrt{p_i^2+m_i^2}
\quad\text{with}\quad
    i=1,2
\enspace .
\end{equation}
We use the three-momentum
conservation to integrate out the momentum $p_1$, i.e. $\vec{p}_{1}=-\vec{p}_{2}$.
This gives us
\begin{equation}
    d\varPi_\mathrm{LIPS}=\frac{1}{\left(2\pi\right)^{2}}\frac{1}{4E_1E_2}d^{3}p_{2}\delta\left(m_{q}-E_1-E_2\right),
\end{equation}
where $m_{q}$ is the mass of the decaying particle. Going to 
spherical coordinates we get
\begin{equation}
    d\varPi_\mathrm{LIPS}
=
    \frac{1}{4\pi}
    \frac{1}{E_1 E_2}
    \left|\vec{p}_{2}\right|^{2}
    d\left|\vec{p}_{2}\right|
    \delta\left(m_{q}-E_1-E_2\right)
    \frac{d^{2}\Omega_{2}}{4\pi}
\enspace .
\end{equation}
Our outgoing particles have the same magnitude of the momentum, but opposite direction:
\begin{equation}
    \left|\vec{p}_{1}\right|=\left|\vec{p}_{2}\right|=p
\enspace .
\end{equation}
Using the energy conservation relation
\begin{equation}
    m_q=E_1+E_2
\enspace ,
\end{equation}
we get that the squared magnitude of the momenta is
\begin{equation}
    p^{2}=\frac{\left(m_{q}^{2}-m_1^{2}-m_2^{2}\right)^{2}-4m_{1}^{2}m_2^{2}}{4m_{q}^{2}}
\enspace ,
\end{equation}
which is symmetric under the interchange $m_1\longleftrightarrow m_2$.

Changing the variable to 
\begin{equation}
    x=E_1+E_2
\enspace ,
\end{equation}
with the differential of $x$
\begin{equation}
    dx=d\left(E_1+E_2\right)=\frac{p}{\sqrt{p^2+m_1^2}}dp+\frac{p}{\sqrt{p^2+m_2^2}}dp=\frac{E_1+E_2}{E_1E_2}pdp 
\enspace , 
\end{equation}
we express $dp$
\begin{equation}
    dp=\frac{E_1E_2}{xp}dx 
\enspace .
\end{equation}
Thus the phase space element becomes
\begin{equation}
    d\varPi_\mathrm{LIPS}
=
    \frac{1}{4\pi}\frac{p}{x}dx\delta\left(m_{q}-x\right) 
    \frac{d^{2}\Omega_{2}}{4\pi}
\enspace.
\end{equation}
In case of no angular dependence of the process we can integrate over the angles, reducing the last fraction to 1. 
Integrating over $x$ gives the final result
\begin{equation}
    \varPi_\mathrm{LIPS}
=
    \frac{1}{4\pi}\frac{p}{m_q}
=
    \frac{\sqrt{\left(m_{q}^{2}-m_1^{2}-m_2^{2}\right)^{2}-4m_{1}^{2}m_2^{2}}}{8\pi m_q^2}
\enspace .
\label{eq:dplipsfinal}
\end{equation}
Some edge cases for this phase space element are given in table~\ref{table_2pLips}.
\begin{table}[h!]
\centering
\begin{tabular}{ |c|c|c|c| } 
 \hline
 & $ m_1=m_2=0$ & $m_1=0\; , \; m_2\neq 0$ & $m_1=m_2=m$\\ 
 \hline
$\varPi_\mathrm{LIPS}$ 
 & \rule[18pt]{0pt}{0pt} $\frac{1}{8\pi}$ \rule[-12pt]{0pt}{0pt}
 & $\frac{1}{4\pi}\frac{m_q^2-m_{2}^2}{2m_q^2}$ 
 & $\frac{1}{8\pi}\sqrt{1 - \frac{4m^2}{m_q^2}}$ \\ 
% & $\frac{1}{8\pi m_q}\sqrt{m_q^2-4m^2}$ \\ 
 \hline
\end{tabular}
\caption{\label{table_2pLips}
Table representing the value of the phase space element in a massless, one massive particle and two particles with the same mass.}
\end{table}
One also has to remember to implement the relations between momenta coming from the phase space also in the squared amplitude.

\section{Expressing \texorpdfstring{$\lambda$}{lambda}s in terms of Higgs masses}
\label{sec:lambdasection}
The parameters of the potential in eq.~\eqref{eq:higgspot}, can also be expressed using the diagonalised mass matrix of the Higgses and the rotation matrix $R$, eq.~\eqref{eq:rhiggsmix}. This is done using the relation~\cite{branco2012}
\begin{equation}
    \widetilde{M}^2=RM^{2}R^{T},
    \label{eq:phitom}
\end{equation}
where $M^{2}=\mathrm{diag}\{m_{h_{0}}^{2},m_{H_{0}}^{2},m_{A}^{2}\}$ is the diagonalised mass matrix and 
\begin{equation}
     \widetilde{M}^2=v^{2}\left(\begin{array}{ccc}
2\lambda_{1} & \mathrm{Re}\left[\lambda_{6}\right] & -\mathrm{Im}\left[\lambda_{6}\right]\\
\mathrm{Re}\left[\lambda_{6}\right] & \frac{m_{h_2^{+}}^{2}}{v^{2}}+\frac{1}{2}\left(\lambda_{4}+2\mathrm{Re}\left[\lambda_{5}\right]\right) & -\mathrm{Im}\left[\lambda_{5}\right]\\
-\mathrm{Im}\left[\lambda_{6}\right] & -\mathrm{Im}\left[\lambda_{5}\right] & \frac{m_{h_2^{+}}^{2}}{v^{2}}+\frac{1}{2}\left(\lambda_{4}-2\mathrm{Re}\left[\lambda_{5}\right]\right)
\end{array}\right)
\end{equation}
is the general mass matrix of the 2HDM, which is derived from the potential in eq.~\eqref{eq:higgspot}. Using the previous relation, eq.~\eqref{eq:phitom}, we can express the $\lambda$s in terms of the diagonalised mass matrix and rotation elements. They are
\begin{equation}
\lambda_{1}=\frac{1}{2v^{2}}\left(RM^{2}R^{T}\right)_{11},\label{eq:l1}
\end{equation}
\begin{equation}
\lambda_{4}=\frac{1}{v^{2}}\left[\left(RM^{2}R^{T}\right)_{22}+\left(RM^{2}R^{T}\right)_{33}-2m_{h_2^{+}}^{2}\right],
\end{equation}
\begin{equation}
\mathrm{Re}\left[\lambda_{5}\right]=\frac{1}{2v^{2}}\left[\left(RM^{2}R^{T}\right)_{22}-\left(RM^{2}R^{T}\right)_{33}\right],
\end{equation}
\begin{equation}
\mathrm{Im}\left[\lambda_{5}\right]=-\frac{1}{v^{2}}\left(RM^{2}R^{T}\right)_{23},
\end{equation}
\begin{equation}
\mathrm{Re}\left[\lambda_{6}\right]=\frac{1}{v^{2}}\left(RM^{2}R^{T}\right)_{11},
\end{equation}
and
\begin{equation}
\mathrm{Im}\left[\lambda_{6}\right]=-\frac{1}{v^{2}}\left(RM^{2}R^{T}\right)_{13}.
\end{equation}

Additionally, from tadpole diagrams one  can express the parameters $\mu_1$, and $\mu_{12}$ in terms of other $\lambda$s~\cite{branco2012}
\begin{equation}
\mu_{1}=-\frac{\lambda_{1}v^{2}}{2},
\end{equation}
and
\begin{equation}
    \mu_{12}=-\frac{\lambda_6v^2}{2}.
\end{equation}
The parameter $\mu_2$ is fixed from the mass matrix as
\begin{equation}
\mu_{2}=\frac{\lambda_{3}v^{2}}{2}-m_{h_2^{+}}^{2}.
\label{eq:mhp}
\end{equation}
That leaves us with $\lambda_{2},\lambda_{3}$, and $\lambda_{7}$
as free parameters.

\printbibliography
\end{document}